\newif\ifsubmit
\submitfalse

\ifsubmit
    \documentclass[format=acmsmall, review=false]{acmart}
\else
    \documentclass[12pt]{article}
\fi

\ifsubmit
    \usepackage{acm-ec-25}
    \usepackage{booktabs} 
    \usepackage[ruled]{algorithm2e} 
    
    \SetAlFnt{\small}
    \SetAlCapFnt{\small}
    \SetAlCapNameFnt{\small}
    \SetAlCapHSkip{0pt}
    \IncMargin{-\parindent}

    \setcitestyle{authoryear}
\else
    \usepackage[utf8]{inputenc}
    \usepackage{fullpage}
    \usepackage[small,bf]{caption}
\fi
\usepackage{amsmath, url, float}

\ifsubmit\else
    \usepackage[style=alphabetic,backend=biber]{biblatex}
    \bibliography{refs}
\fi

\usepackage{graphicx, xcolor}
\usepackage{tikz, pgfplots}
\pgfplotsset{
    compat=1.16,
}
\usetikzlibrary{topaths,intersections,calc}
\usetikzlibrary{positioning,chains,fit,shapes,calc,arrows.meta}
\usepackage{adjustbox}

\let\phi\varphi
\newcommand{\supp}{\mathbf{supp}}

\newcommand{\BEAS}{\begin{eqnarray*}}
    \newcommand{\EEAS}{\end{eqnarray*}}
    \newcommand{\BEA}{\begin{eqnarray}}
    \newcommand{\EEA}{\end{eqnarray}}
    \newcommand{\BEQ}{\begin{equation}}
    \newcommand{\EEQ}{\end{equation}}
    \newcommand{\BIT}{\begin{itemize}}
    \newcommand{\EIT}{\end{itemize}}
    \newcommand{\BNUM}{\begin{enumerate}}
    \newcommand{\ENUM}{\end{enumerate}}
    
    \newcommand{\cf}{{\it cf.}}
    \newcommand{\eg}{{\it e.g.}}
    \newcommand{\ie}{{\it i.e.}}

    \newcommand{\reals}{\mathbf{R}}

    \newcommand{\bmat}[1]{\begin{bmatrix}#1\end{bmatrix}}
    
    \newcommand{\QED}{~~\rule[-1pt]{8pt}{8pt}}\def\qed{\QED}

    \newif\iftodos

\title{Market Clearing with Semi-fungible Assets}
\ifsubmit
    \author{Submission 273}
\else
    \author{
        Theo Diamandis \\
        \texttt{\small tjdiamandis@gmail.com}
        \and 
        Tarun Chitra \\
        \texttt{\small tarun@gauntlet.xyz}
        \and 
        Guillermo Angeris \\
        \texttt{\small gangeris@baincapital.com}
    }
    \date{May 2025}
\fi

\ifsubmit
\begin{abstract}
    As markets have digitized, the number of tradable products has skyrocketed.
    Algorithmically constructed portfolios of these assets now dominate public
    and private markets, resulting in a combinatorial explosion of tradable
    assets. In this paper, we provide a simple means to compute market clearing
    prices for \emph{semi-fungible} assets which have a partial ordering between
    them. Such assets are increasingly found in traditional markets (bonds,
    commodities, ETFs), private markets (private credit, compute markets), and
    in decentralized finance. We formulate the market clearing problem as an
    optimization problem over a directed acyclic graph that represents
    participant preferences. Subsequently, we use convex duality to efficiently
    estimate market clearing prices, which correspond to particular dual
    variables. We then describe dominant strategy incentive compatible payment
    and allocation rules for clearing these markets. We conclude with examples
    of how this framework can construct prices for a variety of algorithmically
    constructed, semi-fungible portfolios of practical importance.
\end{abstract}
\fi

\begin{document}  
\ifsubmit
    \begin{titlepage}

        \maketitle
        
        \vspace{1cm}
        \setcounter{tocdepth}{2} 
        \tableofcontents
        
    \end{titlepage}
\else
    \maketitle
    \begin{abstract}
    As markets have digitized, the number of tradable products has skyrocketed.
    Algorithmically constructed portfolios of these assets now dominate public
    and private markets, resulting in a combinatorial explosion of tradable
    assets. In this paper, we provide a simple means to compute market clearing
    prices for \emph{semi-fungible} assets which have a partial ordering between
    them. Such assets are increasingly found in traditional markets (bonds,
    commodities, ETFs), private markets (private credit, compute markets), and
    in decentralized finance. We formulate the market clearing problem as an
    optimization problem over a directed acyclic graph that represents
    participant preferences. Subsequently, we use convex duality to efficiently
    estimate market clearing prices, which correspond to particular dual
    variables. We then describe dominant strategy incentive compatible payment
    and allocation rules for clearing these markets. We conclude with examples
    of how this framework can construct prices for a variety of algorithmically
    constructed, semi-fungible portfolios of practical importance.
    \end{abstract}
\fi


\section{Introduction}
Many electronically traded assets are non-fungible. In contrast to a company's
common stock, which is all---by construction---fungible, many other assets are
similar but not exactly identical. For a bond, we might be interested in its
yield and credit rating. For a metal, we may care about not only the price at
which it is traded but also the purity level and delivery location. For a
liquid staking token, we might also care about the reputation of the staking
service providers. In all these examples, the items cannot be totally
ordered---in contrast to a company's stock, which can be ordered by price in
some order book. Instead, these items are partially ordered: some are better
than others, some are worse than others, and some are incomparable. We call
these types of assets \emph{semi-fungible}. In this paper, we develop a market
clearing model for divisible, semi-fungible assets, such as those tokenized on
public blockchains.

\paragraph{Partial orders and embeddings.}
Partial orders are one natural interpolation between complete fungibility and
non-fungibility. In particular, fungible assets can generally be easily ordered
based on their price; a buyer always prefers a lower price. For semi-fungible
assets, the asset price is only one of potentially many relevant dimensions.
These assets naturally have some partial orders where some assets may be
preferred to other assets and be further incomparable to others. As we will
see, examples of semi-fungible assets range from bonds and commodities to
online ad placement and some cryptocurrencies.

\subsection{Examples}
We motivate our setting with several examples that also provide helpful mental 
models to ground the mechanisms we subsequently develop.

\paragraph{Bonds.}
Bond traders historically considered each bond individually. More recently,
systemic credit funds and portfolio trading have increasingly taken over as
electronic trading has become more prevalent on these markets. The assets under
management of these funds has more than doubled in the last
year~\cite{bloombergphdsmbasscredit2024}, and portfolio trading---where large
baskets of bonds are traded together---has similarly doubled since the beginning
of 2024~\cite{whittall2023portfolio}. These developments are a result of the
semi-fungibility of bonds: bonds with the same yield and the same credit rating
are, for many purposes, equivalent. In the words of Matt
Levine~\cite{bloombergquantcredit2024}:
\begin{quote}
    Historically, though, for a long time there was a view that each bond is a
    special snowflake and you had to travel to the ends of the earth to get exactly
    the BBB+ bond that you were looking for. 
    [...]
    The modern view is that bonds are largely linear combinations of
    factor exposures so liquidity is just fine.
\end{quote}
In other words, these assets have a number of properties that can be partially ordered.

\paragraph{Semi-fungible commodities.}
Many financial products in commodities markets---even those traded on
exchanges---possess a certain number of differentiating properties. For
example, a given lot of a particular metal, such as nickel, may have specific
purity levels, delivery locations, or some notion of seller reputation. Buyers
may require a certain minimum level or purity from this lot, or may only trust
a subset of sellers in a given market.

\paragraph{Compute markets.}
For a given compute workload, many different types of compute hardware may be
available. For example, a machine learning model could be trained on the latest
GPU (likely the fastest), on an older GPU (likely at a lower throughput), or on
another type of hardware (\eg, a CPU or a TPU). Buyers may have different
preferences for the hardware they would like to use and the relative
price-performance tradeoffs they are willing to make, but ultimately all buyers
are purchasing FLOPs for some time period.

\paragraph{Liquid (re)staking tokens.}
Issuers of yield-bearing assets, such as treasuries or bonds, pay buyers some
rate of interest over time. In return, the buyer accepts the risk that the
seller may default on their payments. As a result, buyers often have a
preference on the reputation of the issuers. In traditional markets, this
reputation is typically assessed by rating agencies such as S\&P or Moody's. A
similar thing has played out in blockchain systems such as Ethereum and Solana.
These systems have recently witnessed the rise of so-called liquid staking
tokens, issued by staking pools such as Lido and Jito, which promise a yield on
their staked assets. The recent rise of re-staking protocols (\eg, Eigenlayer
and Symbiotic) resulted in a proliferation of these yield-bearing assets. Each
of these assets can differ on many dimensions: the yield rate, the reputation
(and therefore risk) of the (re)-staking protocol, the perceived risk of
default (via slashing), and the redemption mechanism, among many other axes.

\paragraph{Collectibles.}
Finally, collectible markets, such as art markets, sneaker markets, or NFT
markets, exhibit similar partially-ordered properties. For example, a buyer may
want a piece by a particular artist, or a piece from a particular collection,
but may not care about the specifics of the piece beyond these properties. In
NFT markets, this type of preference is common, evidenced by the fact that
a large proportion of the collection usually trades close to the floor
price---the lowest available price of any item in the collection.

\subsection{Related work}
We consider the market clearing problem with a
partial ordering induced by divisible, semi-fungible assets. Our work builds on
ideas that have been previously developed in the literature. We note that, because we consider divisible goods, our work differs from the literature on general combinatorial auctions.

\paragraph{Partially-ordered assets.}
Partially-ordered assets have been considered in the literature, often in the
context of indivisible goods.
This type of structure describes relationships for many types of assets:
options on stocks can be (partially) ordered by their strike price and/or maturity~\cite{ait2003nonparametric}; bonds can be ordered by their rating~\cite{jarrow1997markov}; other debt products can be ordered by their seniority~\cite{longstaff2008empirical}; airline seats can be ordered by their class~\cite{brumelle1993airline}; and so on. Other work has considered partially ordered items in more generality, for example, see~\cite{aristodemou2022discrete}. Most of these works are not concerned with market clearing.

\paragraph{Batched exchanges.}
Our market mechanism works by batching together the preferences of many buyers
and clearing the market at regular intervals, akin to the frequent batched 
auctions proposed by~\cite{budish2015high}. We require the buyer preferences to
be utility functions, which may be defined over portfolios of assets. This
structure is similar to the bidding language developed by~\cite{budish2023flow},
which generalizes the orders from~\cite{kyle2017toward}. In contrast to this work,
our work assumes the existance of and uses a partial order of preferences over the assets.

\paragraph{Market clearing.}
The broader literature on market clearing mechanisms is vast. Eisenberg and Gale
proved that, for a market with divisible goods and buyers with linear utilities,
the market equilibrium conditions are the optimality conditions of a particular
convex program~\cite{eisenberg1959consensus}. The case with concave utilities
maps to a convex optimization problem~\cite{vegh2014concave}. Many results in
the literature have been developed for markets fully fungible or non-fungible
goods using these models frameworks.
(See~\cite{vazirani2007combinatorial,codenotti2007computation} and citations
therein.) A smaller set of work has considered market clearing when participants have preferences over the items. (See, for example,~\cite{chen2011market, gul1999walrasian}.) We focus on the specific case where there exists some global partial ordering over the items in this market, and buyers can express their preferences over the items in terms of this ordering.

\subsection{This paper}
In this paper, we introduce a market clearing model for divisible, semi-fungible assets by
leveraging partial order relationships among different properties. We introduce
a framework that captures partially ordered preferences across items in
\S\ref{sec:asset}. Next, we define a convex optimization formulation for market
clearing in \S\ref{sec:exchange}. In~\S\ref{sec:mechanism}, we derive market
clearing prices and also propose a dominant strategy incentive compatible
mechanism for clearing these markets. We conclude with both toy and real-world
examples in~\S\ref{sec:examples}.

\section{The asset}
\label{sec:asset}
We consider a single asset (\eg, some particular metal, such as nickel) with
some number of \emph{properties}. For example, in metals markets, these
properties may include the metal purity levels, delivery locations, or some
notion of seller reputation. We denote the set of properties (which is some
abstract set) as $\mathcal P$. We will refer to an \emph{item} $i$ as a
(generally divisible) good with property $p_i$ that can be purchased in some
quantity.

\paragraph{Ordering over properties.} 
We assume that these properties are \emph{partially ordered}. That is, given
two items with some properties $p, p' \in \mathcal P$, respectively, we have
one of three cases: either $p \le p'$; or $p \ge p'$; or, if neither of these
are true, then $p$ is \emph{incomparable} to $p'$. The first, $p \le p'$, may
be read as or `$p$ is no better than $p'$', and similarly for the second, after
interchanging $p$ and $p'$. In addition, if there is a third item with
properties $p'' \in \mathcal P$ which satisfy $p'' \ge p'$ then, if $p' \ge p$,
we should have $p'' \ge p$; \ie, the relation is transitive. Finally, we assume
that the set of orders has the property that, if $p \ge p'$ and $p \le p'$,
then $p = p'$; \ie, the relation is antisymmetric.

\paragraph{Interpretations.} 
The ordering over items has the following interpretation. If we have two items
$i$ and $i'$ with properties $p_i$ and $p_{i'}$, respectively, then, if $p_i
\le p_{i'}$, we assume that, all else being equal, anyone would accept the
item with properties $p_{i'}$ if they would accept the items with properties
$p_i$. For example, given two metals, one with higher purity than the other, a
buyer attempting to purchase the lower purity metal would also happily accept
the higher purity metal at the same price, all else being equal. Similarly, a
bond buyer would happily accept a higher yield, all else being equal. The case
with $p_i \ge p_{i'}$ is similar, replacing the role of $i$ with $i'$.
Finally the case where $p_i$ is incomparable to $p_{i'}$ means that some
buyers might prefer item $i$ over $i'$ and some others might prefer $i'$
over $i$. For example, if some metal originates from the United States versus
the Philippines, some buyers might prefer the former over the latter and vice
versa. Similarly, if some bond has a duration of one year and another has a
duration of two years, some buyers might prefer the former over the latter and
vice versa.

\paragraph{Discussion.} 
We will show that it is possible to construct markets that make use of this
additional partial-ordering structure in order to clear. These markets have a
number of important guarantees about how orders are executed. For example, in
this market, if a buyer expresses that she wants an item with properties at
least as good as $p' \in \mathcal P$, then, if no such item is available, she
can be matched with any item $i$ that has properties $p_{i} \ge p'$, assuming
such a item exists and is provided at no more than the price she is willing to
pay. (In economics parlance, we say that she has access to `more liquidity' in
this market than a market that only trades items with properties $p_i$.)

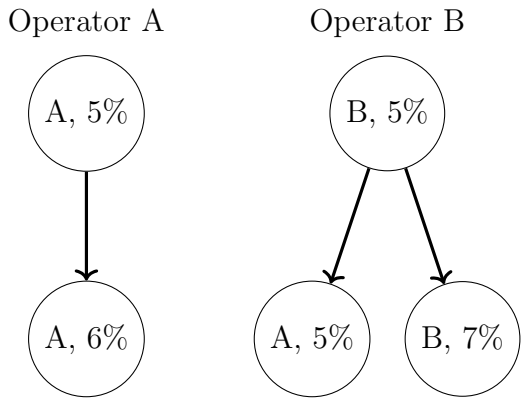
\begin{figure}[h]
    \centering
    \adjustbox{max width=0.48\textwidth}{
    \begin{tikzpicture}
    
    \node[draw, circle] (A1) at (0, 0) {A, 5\%};
    \node[draw, circle] (A2) at (0, -3) {A, 6\%};
    
    \node[draw, circle] (B1) at (4, 0) {B, 5\%};
    \node[draw, circle] (A3) at (3, -3) {A, 5\%};
    \node[draw, circle] (B2) at (5, -3) {B, 7\%};
    
    \draw[->, very thick] (A1) -- (A2);
    \draw[->, very thick] (B1) -- (A3);
    \draw[->, very thick] (B1) -- (B2);
    
    \node at (0, 1.2) {Operator A};
    \node at (4, 1.2) {Operator B};
    
    \end{tikzpicture}
    }
    \caption{An example partial ordering on five yield-bearing assets with
    different ratings and yields, provided by different operators. We assume
    the operators are not comparable, whereas any buyer prefers a higher rating
    and higher yield, all else being equal.}
    \label{fig:partial-order-dag}
\end{figure}

\paragraph{Partial orders as DAGs.} 
Any partial order with a finite number of elements can be viewed as a directed
acyclic graph (often abbreviated to DAG) with $P = |\mathcal P|$ vertices, which we
will label with the elements of $\mathcal P$. The edges are defined in the
following way: the edge from $p \in \mathcal P$ to $p' \in \mathcal P$ is
present if $p \le p'$ and there is no $p'' \in \mathcal P$ such that $p \le p''
\le p'$. In other words, two nodes have an edge between them if there is no
element that lies in between them (with respect to the ordering of $\mathcal
P$). We will use this representation in what follows. A simple example for
yield-bearing assets is shown in figure~\ref{fig:partial-order-dag}.

\begin{figure}
    \centering
    \adjustbox{max width=0.48\textwidth}{
        \begin{tikzpicture}[every node/.style={inner sep=2pt}, thick]
            \node (A) at (0,-2) {$\{A\}$};
            \node (B) at (2,-2) {$\{B\}$};
            \node (C) at (4,-2) {$\{C\}$};
        
            \node (AB) at (0,0) {$\{A, B\}$};
            \node (AC) at (2,0) {$\{A, C\}$};
            \node (BC) at (4,0) {$\{B, C\}$};
        
            \node (ABC) at (2,2) {$\{A, B, C\}$};
        
            \draw[<-] (A) -- (AB);
            \draw[<-] (A) -- (AC);
            \draw[<-] (B) -- (AB);
            \draw[<-] (B) -- (BC);
            \draw[<-] (C) -- (AC);
            \draw[<-] (C) -- (BC);
        
            \draw[<-] (AB) -- (ABC);
            \draw[<-] (AC) -- (ABC);
            \draw[<-] (BC) -- (ABC);
        \end{tikzpicture}    
    }
    \caption{Partial orders on the set of, say, node operators. Some users may 
    be indifferent between the operators $A$ and $B$, while others may be 
    indifferent between $A$ and $C$, and so on.}
\end{figure}
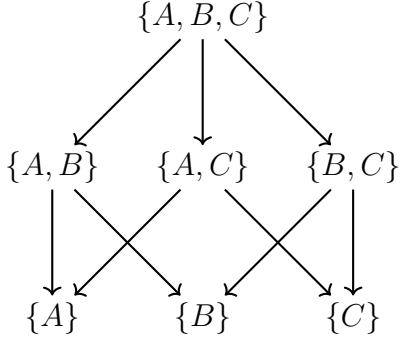

\subsection{The embedding}\label{sec:embedding}
Here, we discuss how to express buyer preferences for the items.

\paragraph{Embedding the ordering.}
A buyer $b$ can express her preferences for the asset by specifying a property
$p_b \in \mathcal P$ which will, in some sense, be the base item she is willing
to purchase. In other words, given a list of $n$ items $i=1, \dots, n$ with
properties $\{p_i\}$, she will take any item $i$ satisfying $p_i \ge p_b$.

\paragraph{Preference vector.}
Let the vector $z_b \in \reals^n_+$ denote the amount of each of the $n$ items
allocated to buyer $b$. (Buyer $b$ need not want all of these items.) We then
write $x_b \in \reals_+$ for the total amount of the asset (which includes any
items with properties at least as good as $p_b$) that buyer $b$ is willing to
buy. Written out in math, this is
\[
    x_b = \sum_{i\,: \,p_i \ge p_b} (z_b)_i.
\]
The sum is taken over all items $i$ such that with properties $p_i$ are at least
as good as buyer $b$'s desired property; \ie, the possible items' properties
satisfy $p_i \ge p_b$. This can be written much more succinctly as
\[
    x_b = c_b^Tz,
\]
where we call the vector $c_b \in \{0, 1\}^n$ the \emph{preference vector}
for buyer $b$, defined
\[
    c_b = \sum_{i\,: \,p_i \ge p_b} e_i.
\]
(Here, $e_i$ it the $i$th unit basis vector.) More generally, we may wish to
express weighted preferences between the categories. For example, if a buyer
has access to a zero-cost purification process, she may be indifferent between
one unit of an 100\% pure metal and two units of a 50\% pure metal. We can
express this preference using a vector $c \in [0,1]^n$, such that the `total
amount' of the asset received is $c^T z$. We leave these more general,
non-Boolean preference vectors as an extension.

\paragraph{Preference vector ordering.}
Note that the preference vectors of buyers are themselves also ordered in the
following way: for any buyers $b$ and $b'$, we have
\[
    p_b \le p_{b'} \quad \text{if, and only if,} \quad c_b \ge c_{b'},
\]
where the first inequality is with respect to the partial ordering over
$\mathcal P$ and the second is an elementwise inequality. Indeed, letting
\[
    \supp(z) = \{i \mid z_i \ne 0\},
\]
be the \emph{support} of a vector $z$; \ie, the set of indices for which $z$ is
nonzero, then this is equivalent to saying that 
\begin{equation}
    \label{eq:embedding}
    p_b \le p_{b'} \quad \text{if, and only if,} \quad \supp(c_b) \supseteq \supp(c_{b'}),
\end{equation}
since $c_b$ and $c_{b'}$ are Boolean. In other words, buyers willing to accept
items with worse properties have more permissive preference vectors. We make use 
of this relationship in what follows.

\section{The exchange}
\label{sec:exchange}
We now specify the market clearing problem for partially ordered assets, as
described above. We consider a single asset with $B$ buyers, labeled $b = 1,
\dots, B$ and one supplier with a quantity $q \in \reals^n_+$ of each item to
sell. The $i$th entry of $q$, denoted $q_i$, represents the total quantity of
item $i$ (with property $p_i$) that the supplier is willing to sell. Each buyer
$b$ receives some amount $x_b \in \reals_+$ of the asset after the market
clears. 

\paragraph{Preferences.}
As discussed above, we use a vector $c_b \in \reals^{n}$ to represent the
preferences of a buyer over the items. If the buyer expresses that she wants
items with properties $p' \in \mathcal P$, then, if no such item is available,
she can be matched with any item $i$ that has properties $p_i \ge p'$. Thus,
$(c_b)_i = 1$ for all $i$ such that $p_i \ge p'$ and $(c_b)_i = 0$ otherwise.

\paragraph{Allocations.}
We define vectors $z_{b} \in \reals^n_+$ such that the $i$th entry of $z_{b}$,
denoted $(z_{b})_i$, is the amount of the item with property $p_i$ that the
supplier sells to buyer $b$. The total amount of the asset received by buyer
$b$ (that this buyer is willing to take) can be written as
\[
    x_b = c_b^T z_{b}.
\]
Finally, we include a constraint ensuring the supplier does not sell more than
the available items:
\[
    \sum_{b=1}^B z_{b} \le q.
\]

\paragraph{Utilities.}
We collect the amount of the asset received by each buyer into a vector $x =
(x_1, \dots, x_B) \in \reals^B_+$. We express the utility of this allocation by
a concave, nondecreasing function $U : \reals^B_+ \to \reals \cup \{-\infty\}$,
where infinite values encode constraints: an allocation $x$ such that $U(x) =
-\infty$ is unacceptable. We also require that $U(0) = 0$ and assume that $U$ is
strictly increasing at $0$, \ie, if $U$ is differentiable at $0$, that $\nabla
U(0) > 0$. We note that $U$ is typically separable over the buyers; each buyer
individually specifies some utility function on their individual allocation,
denoted $u_b: \reals_+ \to \reals \cup \{-\infty\}$. In this case, we can write
$U$ as
\[
    U(x) = \sum_{b = 1}^B u_b(x_b).
\]

\paragraph{Market clearing.}
We write the market clearing problem as
\begin{equation}
    \label{eq:market-clearing}
    \begin{aligned}
        &\text{maximize} && U(x) \\
        &\text{subject to} && x_b = c_b^T z_{b} \\
        &&& \sum_{b=1}^B z_{b} \le q\\
        &&& z_b \ge 0, \qquad b = 1, \dots, B.
    \end{aligned}
\end{equation}
The variables are $x \in \reals^B$ and $z_b \in \reals^n$ for each buyer $b=1,
\dots, B$. The problem data are the items supplied to the market $g \in
\reals^n_+$, the preferences $c_b \in \reals^n_+$, and the utility function $U$,
which is often a sum of buyer utilities $u_b$. This problem is easily recognized
as a convex optimization problem with linear constraints. In fact, when the
utility function $U$ is piecewise linear, this problem is simply a linear
program. In both cases, the problem can be solved efficiently in practice. Note
that this problem can also be written as a special case of a network flow
problem in the framework of~\cite{diamandis2024convex} (\cf, the Fisher market
problem introduced in~\cite[\S3.4]{diamandis2024convex}).


\paragraph{Discussion.}
This market model also extends to the common scenario where buyers may wish to
purchase multiple `baskets' of items, each corresponding to different
preferences and utility functions. For example, a bond buyer may want to
express preferences over higher-rated and lower-rated bonds separately. In this
case, we can simply introduce two `buyers' $b$ and $b'$ corresponding to these
different preferences in our model. Since the utility function is concave, this
model allows for preferences with complements. (This is the case, for example,
if we value two units of items which have properties at least as good as both
$A$ and $B$ more than we value one unit of items with properties at least as
good as $A$ and one unit of items at least as good as $B$, but not necessarily
both.) Our model and the associated analysis can also be easily extended to the
scenario in which we have more than one seller, each with some allocation of
items, and to the scenario with more than one asset.

\section{Market-clearing mechanism}
\label{sec:mechanism}
Now that we have defined the market clearing problem~\eqref{eq:market-clearing},
which is a convex optimization problem, we derive and examine the dual problem
to determine how we should clear this market.

\subsection{Market-clearing prices}
\label{subsec:market-clearing-prices}
We can write the Lagrangian of the market clearing problem as
\[
    L(x, z, \nu, \lambda) = 
        U(x) 
        + \sum_{b=1}^B \nu_b(c_b^Tz_b - x_b) 
        + \lambda^T\left(q - \sum_{b=1}^B z_b\right) 
        - \sum_{b=1}^B I(z_b),
\]
where the dual variables are $\nu \in \reals^B$ and $\lambda \in \reals^n_+$,
while $I: \reals^n \to \reals \cup \{+\infty\}$ is the indicator function for
the nonnegative reals, defined as
\[
    I(w) = 
    \begin{cases}
        0 & w \ge 0\\
        \infty & \text{otherwise}.
    \end{cases}
\]
Maximizing over the primal variables $x$ and $\{z_b\}$, we obtain the dual
function
\[
    g(\nu, \lambda) 
    = \lambda^T q
    + \sup_x\left( U(x) - \nu^T x \right) 
    + \sum_{b=1}^B \sup_{z_b \ge 0} \left((\nu_b c_b - \lambda)^Tz_b\right)
\]
The dual problem is then
\begin{equation}
    \label{eq:dual-problem-impl}
    \begin{aligned}
        &\text{minimize} && g(\nu, \lambda) \\
        &\text{subject to} && \lambda \ge 0,
    \end{aligned}
\end{equation}
with variables $\lambda \in \reals^n$ and $\nu \in \reals^B$. Note that this
problem is also convex, as $g$ is the pointwise supremum of a family of affine
functions.

\paragraph{Duality.}
Denote the optimal value of the primal problem~\eqref{eq:market-clearing} by
$p^*$ and the optimal value of the dual problem~\eqref{eq:dual-problem-impl} by
$d^*$. Since all constraints are linear and the problem has nonempty feasible
set, we have strong duality~\cite[Prop.\ 5.3.1]{bertsekas2009convex}:
\[
    p^* = d^*,
\]
and there exists some dual variables $\lambda^*$ and $\nu^*$ which achieve this
bound.

\paragraph{Implicit constraints.}
Problem~\eqref{eq:dual-problem-impl} has some additional implicit constraints.
Observe that, if there exists some index $i$ and buyer $b$ such that 
$(\nu_b c_b - \lambda)_i > 0$, then 
\[
    \sup_{z_b \ge 0} \left((\nu_b c_b - \lambda)^Tz_b\right) 
    \ge (\nu_b c_b - \lambda)^T (t e_i) = t(\nu_b c_b - \lambda)_i \to \infty,
\]
as $t \uparrow \infty$. Here, we have chosen $z_b = t e_i$, where $e_i$ is the
$i$th unit basis vector. This implies that the supremum is infinite if any
element of $\nu_b c_b - \lambda$ is positive, so we must have
\[
    \nu_b c_b - \lambda \le 0
\] 
as an implicit constraint for each $b = 1, \dots, B$. By a similar argument,
we can show that if $U$ is nondecreasing and $U(0) < \infty$, then we have the
implicit constraint that
\[
    \nu \ge 0.
\]
Adding both implicit constraints as explicit
constraints gives the following constrained optimization problem:
\begin{equation}
\label{eq:dual-problem}
\begin{aligned}
    &\text{minimize} && \sup_x\left( U(x) - \nu^T x \right)  + \lambda^T q \\
    &\text{subject to} && \nu_b c_b \le \lambda, \quad b = 1, \dots, B \\
    &&& \lambda \ge 0,\; \nu \ge 0.
\end{aligned}
\end{equation}

\paragraph{Simplification.}
Note that the first constraint tells us that $\lambda$ dominates $\nu_b c_b$
for each buyer $b=1, \dots, B$. Using this, and the fact that $q \ge 0$, we can
further simplify the dual problem:
\begin{equation}
\label{eq:dual-problem-simplified}
\begin{aligned}
    &\text{minimize} && \sup_x\left( U(x) - \nu^T x \right)  
    + \sum_{i=1}^n q_i \max_{b} ((\nu_b c_b)_i) \\
    &\text{subject to} && \nu \ge 0.
\end{aligned}
\end{equation}
After this simplification, the dual problem only has a positivity constraint on
the dual variables and is relatively easy to solve with standard methods. This
form also suggests that a tatonnement-style algorithm could be used to solve
this problem, although we believe that solving the original 
problem~\eqref{eq:market-clearing} makes more sense in practice.

\paragraph{Dual variables.}
Both dual variables $\nu$ and $\lambda$ have natural interpretations. To
simplify the exposition, we will let $U$ be differentiable in what follows, but
note that this is not necessary and a suitable generalization exists via
subgradients. From the supremum in the objective of~\eqref{eq:dual-problem}
and~\eqref{eq:dual-problem-simplified}, we have that the optimal dual variable
$\nu^\star$ and optimal allocation $x^\star$ will satisfy
\[
    \nabla U(x^\star) = \nu^\star.
\]
Thus, the dual variable $\nu^\star$ gives the marginal utilities for each buyer
at the optimal allocations. Viewing these dual variables as prices, we
interpret the objective in~\eqref{eq:dual-problem-simplified} as maximizing the
net utility (utility minus cost) of the buyers, plus the seller's revenue from
each item. (Recall that $(c_b)_i = 1$ if buyer $b$ would accept item $i$ with
properties $p_i$ and $0$ otherwise.) While we may view an optimal dual variable
$\nu^\star$ as the marginal prices for each buyer, we may view an optimal dual
variable $\lambda^\star$ as the marginal price for each item. In particular, we
know that at optimality, for the same reason we may
rewrite~\eqref{eq:dual-problem} as~\eqref{eq:dual-problem-simplified}, we have
\begin{equation}\label{eq:dual-variable-relation}
    \lambda_i^\star = \max_b ((\nu_b c_b)_i),
\end{equation}
for each $i=1, \dots, n$. (This is an optimal point if $q \ge 0$, but not
necessarily uniquely so.) Since the buyers themselves are partially sorted by
preferences (from~\eqref{eq:embedding}), this observation says that that, over
all buyers $b$ willing to buy item $i$, $\lambda_i^\star$ is the maximum
marginal price that any of these buyers is willing to pay for this item. In
turn, this implies that the marginal price for each item $i$ respects the
partial ordering of properties in the sense that
\[
    \lambda^\star_i \ge \lambda^\star_{i'} \quad \text{if} \quad p_i \ge p_{i'}.
\]
In English: if any item $i$ has properties at least as good as another item
$i'$, then the marginal price of $i$ is at least as large as that of $i'$. (This
makes sense; any buyer of item $i'$ would always prefer to buy item $i$ for any
price below that of $i'$, since item $i$ is at least as good.) Note that this
relationship does not say anything about the price of items that are
incomparable.

\paragraph{Implementation and payments.}
We can implement this mechanism by asking users to specify their preferences $U$
and then solving the allocation problem~\eqref{eq:market-clearing} for the
optimal allocation $x^\star$ and dual variable $\nu^\star$. We can charge each
user $i$ their marginal utility $\nu^\star_i$ for their allocation. In this
case, this mechanism is essentially a first-price auction over an infinite set
of goods. As in a standard first-price auction, we would expect users to shade
their preferences in order to maximize their net utility (utility minus cost).
In the following section, we take advantage of the convex nature of this market
clearing problem to construct and incentive compatible payment rule.

\subsection{A DSIC mechanism}\label{sec:dsic}
While duality provides market-clearing prices $\nu$ for each buyer, these prices
are not necessarily incentive-compatible. Thus, the question remains: if we
clear the market by solving~\eqref{eq:market-clearing}, what do we charge each
buyer for their allocation? We answer this question by viewing the market
clearing problem as a system welfare computation. This viewpoint leads to an
`auction' allocation rule and the associated dominant strategy incentive
compatible (DSIC) payment rule. 

\paragraph{Max-of-sum welfare.} In this construction, we will have some
compact (under whatever topology) space of allocations $S$ and some
locally-convex topological vector spaces $\mathcal{F}_1, \dots, \mathcal{F}_B$
of upper semicontinuous functions (with respect to the topology of $S$) mapping
$S \to \reals$, potentially with some additional restrictions such as concavity
or monotonicity. We will define the following family, indexed by the set $S$
and functional spaces $\mathcal{F} = \mathcal{F}_1 \times \dots \times
\mathcal{F}_B$, of welfare functions $F: \mathcal{F} \to \reals$ of the form:
\begin{equation}\label{eq:max-of-sep}
    F(f) = \sup_{x \in S} \left(\sum_{b=1}^B f_b(x)\right),
\end{equation}
where $f\in \mathcal{F}$. We call these types of functions $F$
\emph{max-of-sum} due to their form. In this case, we can interpret the
function $f_b\in \mathcal{F}_b$ as the \emph{bid} for buyer $b$ (who gets to
construct any function they wish, as in~\eqref{eq:market-clearing}) and $S$ as
the set of possible allocations (the constraints of~\eqref{eq:market-clearing}.
If all buyers truthfully report their bid, this $F$ would correspond exactly
to the welfare gained by buyers and any solution $x^\star \in S$ (which exists
by compactness of $S$ and upper-semicontinuity of the $\{f_i\}$) to the problem
is a welfare-maximizing allocation. It is not, of course, a-priori obvious that
this will be the case, or how much the mechanism should charge users in order
to ensure truthful bidding. We deal with these issues next.

\paragraph{Allocation rules.} 
We define an \emph{allocation rule}
as a probability measure $\mu$ over (a sigma algebra of subsets of) the set
$S$, for any fixed vector of bids $f$. (In our particular case, we will have $S
\subseteq \reals^n$, which means that the usual sigma-algebra over $\reals^n$
works, but this construction holds more generally.) Any allocation is a sample
drawn from the distribution $\mu$. In our particular case, the allocation rule
will be very simple: set $\mu(\{x^\star\})=1$, where $x^\star$ is any solution
to the right hand side of~\eqref{eq:max-of-sep}. In other words, we set the
allocation to always be a point mass on any solution of the right hand side
of~\eqref{eq:max-of-sep}. Note that any probability distribution
with support over the optimal points would also work. These rules might be of
interest to ensure certain types of `fairness' when the solution is not unique.

\paragraph{Myerson's lemma.} Myerson's lemma~\cite{myerson1981optimal} gives us
a way of constructing DSIC payment rules for allocation rules that are monotone
in the finite-dimensional setting (\ie, when $\mathcal{F}$ is itself a finite
dimensional vector space over $\reals$). Unfortunately, it is not clear what the
correct generalization is when the space of possible bids $\mathcal{F}$ need not
be finite-dimensional, as is true in the partially ordered market construction
of~\S\ref{sec:exchange}. We will explicitly construct a payment rule and then
show that it is DSIC under quasilinear utilities, although the result holds in
more general settings.

\paragraph{Payment rule.} The payment $P_b(f)$ that buyer $b$ must make for a
given set of bids $f \in \mathcal{F}$ and allocation rule (distribution) $\mu$,
is
\[
    P_b(f) = F(f_{-b}) - \sum_{b'\ne b}\int_S f_{b'}\,d\mu,
\]
where $f_{-b} = (f_1, \dots, f_{b-1}, 0, f_{b+1}, \dots, f_B) \in
\mathcal{F}$. In the case that we choose the allocation rule $\mu$ to always
be a point mass at the optimal point $x^\star$ which
maximizes~\eqref{eq:max-of-sep} over the bids $f$, this simplifies to
\begin{equation}\label{eq:payment-rule}
    P_b(f) = F(f_{-b}) - \sum_{b'\ne b}f_{b'}(x^\star).
\end{equation}
Note that this payment rule can be evaluated by solving the market clearing
problem~\eqref{eq:market-clearing} $B+1$ times: once to compute an optimal
allocation $x^\star$, and $B$ times to compute the allocation without each of
the $B$ buyers.

\paragraph{Interpretation.} The interpretation of this rule is particularly
simple: we charge buyer $b$ the marginal utility lost by all other buyers for
having included buyer $b$'s bid. In other words, buyer $b$ must pay their
externality. This payment rule uses essentially the same idea as the Vickrey
auction, except in this more general case. We can recover the Vickrey auction
payment and allocation rules directly from this construction as the special
case where $S = \{e_b \mid b=1, \dots, B\}$ where $e_b$ is the $b$th unit basis
vector and the functions $f_b$ are linear and depend only on the $b$th
component; \ie, when $f_b: S \to \reals$ and the family of possible functions
are those for which 
\[
    \mathcal{F}_b = \{f_b \mid f_b(x)= \alpha_bx_b ~\text{for some $\alpha_b \in \reals_+$} \},
\]
for each $b=1, \dots, B$. For the Vickrey auction with a single item, the set
$S$ simply says that the item must be allocated to one buyer and
problem~\eqref{eq:max-of-sep} says that the welfare is the allocation which
takes the maximum over all bids. The payment rule~\eqref{eq:payment-rule} then
says that the winning buyer (ties broken based on the definition of $\mu$) pays
the second-highest bid and the rest pay nothing. We leave extensions of this
model to handle reserve prices, indivisible goods, and other variants considered
in the literature to future work. We also note that the usual issues with sybils
in VCG mechanisms also apply here (false-name bids)~\cite{yokoo2004effect}. We leave exploration of false-name-proof mechanisms to future work.

\paragraph{Proof of DSIC.} The proof that this mechanism is dominant strategy
incentive compatible is mostly notation-chasing. Let $\bar f_b: \mathcal{F}_b
\to \reals$ be any bid for buyer $b$ and let $f_b: \mathcal{F}_b \to \reals$
be the true valuation; \ie, for any allocation $\mu$, the (expected) value to
player $b$ of this allocation is $\int_S f_b\,d\mu$. Then, by definition,
$f_{-b}$ does not contain player $b$'s bid (given by $\bar f_b$). We will show
that the payoff for player $b$ having bid $f_b$ is always at least as large as
the payoff for her having bid any other $\bar f_b$. In particular, consider the
difference in payoffs:
\[
\underbrace{\left(
    \int_S f_b\,d\mu - \left(F(f_{-b}) - \sum_{b'\ne b} \int_\mu f_{b'}\,d\mu
\right)\right)}_\text{payoff for bidding $f_b$} 
- \underbrace{\left(
    \int_S f_b\,d\bar \mu - \left(F(f_{-b}) - \sum_{b'\ne b} \int_\mu f_{b'}\,d\bar\mu
\right)\right)}_\text{payoff for bidding $\bar f_b$}.
\]
Here $\mu$ denotes any allocation rule consistent with our definition (that is,
any probability distribution over maximizers of the right hand side
of~\eqref{eq:max-of-sep}) for the bids $(f_1, \dots, f_B)$ and $\bar \mu$ is the
same but for the bids $(f_1, \dots, f_{b-1}, \bar f_b, f_{b+1}, \dots f_B)$.
Rearranging gives that this expression is equal to
\[
    \sum_{b=1}^B\int_S f_b\,d\mu - \sum_{b=1}^B\int_S f_b\,d\bar\mu 
    = F(f) - \int_S \left(\sum_{b=1}^B f_b\right)\,d\bar \mu \ge 0.
\]
Where the inequality follows from the fact that any expectation is never larger
than the supremum, along with the definition of $F$.

\section{Examples}
\label{sec:examples} 
We conclude this paper with a few examples. First, we show a simple numerical
example to demonstrate the payment and allocation rules, and how prices change
under different preferences. Next, we show how to model decentralized lending
protocols, which offer a variety of different yield products, in our framework.
Finally, we model restaking networks, in which the utility for any buyer depends
not only on their allocation but also on the actions of all other buyers. We
show that this situation fits into our market clearing
mechanism~\S\ref{sec:exchange} and conjecture that the DSIC payment rule
developed in~\S\ref{sec:mechanism} extends as well.

\subsection{Simple example}
\label{sec:simple-ex}
\begin{figure}[h]
    \centering
    \adjustbox{max width=0.48\textwidth}{
    \begin{tikzpicture}
    
    \node[draw, circle] (B1) at (4, 0) {B, 5\%};
    \node[draw, circle] (B3) at (3, -3) {A, 6\%};
    \node[draw, circle] (B2) at (5, -3) {B, 7\%};
    
    \draw[->, very thick] (B1) -- (B3);
    \draw[->, very thick] (B1) -- (B2);
    
    \end{tikzpicture}
    }
    \caption{ Three yield bearing assets with various ratings, and their
    associated partial ordering. }
    \label{fig:partial-order-example}
\end{figure}
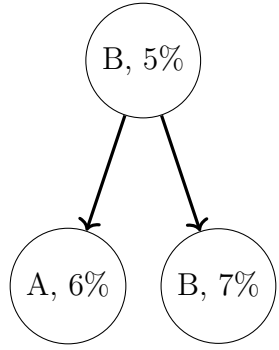

Consider a market with $B=2$ buyers and $n = 3$ yield bearing assets with
different ratings and yields, depicted in
figure~\ref{fig:partial-order-example}.
There is one (divisible) unit of each item. We consider a few different
scenarios and look at the market clearing prices (dual variables) and the
payment rules in each case.

\paragraph{Homogenous market.}
First, consider the case where each buyer would accept any item. Buyers express
utilities in units of yield. The resulting (identical) preference vectors are 
\[
\begin{aligned}
    c_1 = c_2 = (6, 5, 7)
\end{aligned}
\]
where assets are indexed via a post-order traversal of the graph in
figure~\ref{fig:partial-order-example}. We assume that both buyers have a square
root utility function
\[
    u_b(x_b) = \sqrt{x_b}.
\]
Solving the market clearing problem~\eqref{eq:market-clearing} results in 
symmetric allocation and payments:
\[
    x_1^\star = x_2^\star = 9, 
    \qquad \text{and} \qquad 
    z_1^\star = z_2^\star = \bmat{0.5\\ 0.5\\ 0.5}.
\]
The dual variables are $\nu_1^\star = \nu_2^\star = 1/6$, and we can easily
verify that~\eqref{eq:dual-variable-relation} holds:
\[
    \nu_1^\star \cdot c_1 
    = \nu_2^\star \cdot c_2 
    = \frac{1}{6} \cdot \bmat{6\\ 5\\ 7} = \lambda^\star = \bmat{1\\ 5/6\\ 7/6}.
\]
Each buyer pays $1.24$ resulting in a net utility of $1.76$.

\paragraph{Rating preferences.}
Next, we consider the case where buyer $1$ only wants an A-rated asset, but
buyer $2$ is indifferent.  The following vectors encode these preferences:
\[
\begin{aligned}
    c_1 &= (6, 0, 0)\\
    c_2 &= (6, 5, 7)
\end{aligned}
\]
Both buyers again have a square root utility function. Solving the market
clearing problem~\eqref{eq:market-clearing} results in the allocation
\[
    x_1^\star = 6, \quad z_1^\star = \bmat{1 \\ 0 \\ 0}
    \qquad \text{and} \qquad 
    x_2^\star = 12, \quad z_1^\star = \bmat{0 \\ 1 \\ 1}.
\]
The dual variables are $\nu_1^\star = 0.204$. $\nu_2^\star = 0.144$, and we can
easily verify that~\eqref{eq:dual-variable-relation} holds:
\[
    \nu_1^\star \cdot c_1 =  \bmat{1.224 \\ 0 \\ 0}
    \le \lambda^\star = \bmat{1.225 \\0.722 \\1.01}
    \qquad \text{and} \qquad
    \nu_2^\star \cdot c_2 =  \bmat{0.864 \\0.72 \\1.01} 
    \le \lambda^\star = \bmat{1.225 \\0.722 \\1.01},
\]
with each elementwise inequality saturated for at least one buyer.
In this case, buyer $1$ pays $0.78$ but buyer $2$ pays nothing. This payment 
makes sense; buyer $2$ was only allocated items that buyer $1$ would not accept.
The resulting net utilities are $1.67$ and $3.47$ respectively.

\paragraph{Different utilities.}
Finally, we consider the case where the buyers have different utility functions
but both are willing to accept either A- or B-rated assets. Here, we use the
utility functions
\[
    u_1(x_1) = \log(x_1 + 1) \qquad \text{and} \qquad u_2(x_2) = \sqrt{x_2}.
\]
Solving the market clearing problem~\eqref{eq:market-clearing} results in the
allocation
\[
    x_1^\star = 5.94, \quad z_1^\star = \bmat{0.34\\ 0.35 \\ 0.31}
    \qquad \text{and} \qquad 
    x_2^\star = 12.06, \quad z_1^\star = \bmat{0.66 \\ 0.65 \\ 0.67}.
\]
The dual variables are $\nu_1^\star = 0.144$. $\nu_2^\star = 0.144$, and we can
easily verify that~\eqref{eq:dual-variable-relation} holds:
\[
    \nu_1^\star \cdot c_1 =  \bmat{0.864 \\0.72\\1.01}
    \le \lambda^\star = \bmat{0.864\\ 0.72\\ 1.008}
    \qquad \text{and} \qquad
    \nu_2^\star \cdot c_2 =  \bmat{0.864 \\0.72 \\1.01} 
    \le \lambda^\star = \bmat{0.864\\ 0.72\\ 1.008},
\]
This result makes sense; the buyers have different utility functions, but the
market clearing mechanism chose an allocation such that their marginal utilities
match. In this case, buyer $1$ pays $0.77$ and buyer $2$ pays $1.01$. The
resulting net utilities are $5.17$ and $2.47$ respectively.

\subsection{Decentralized lending protocols}
On-chain lending markets match suppliers, who want to earn yield on their
existing holdings, to users who want to borrow against their existing portfolio.
These protocols are overcollateralized: borrowers have to deposit collateral
greater than the initial value of their loan to open a position. Such loans,
which are similar to securities-based lending and Lombard loans, allow a user
who holds a large quantity of a risky asset (such as Ethereum) to borrow
num\'eraire assets (such as the dollar-pegged USDC) to invest elsewhere. These
markets have grown to over \$35 billion in supplied assets and have had at least
\$2 billion in daily outstanding loans since 2021~\cite{defillama-aave}.

\paragraph{Classic DeFi lending protocols.}
The first on-chain lending protocols within decentralized finance (DeFi) include
Compound~\cite{kao2020analysis} and Aave~\cite{fuAave,sun2022liquidity}. These
protocols pool together suppliers' assets and algorithmically set borrow
interest rates. Losses from borrowers who default are shared amongst the pool,
reducing the variance in returns for suppliers. However, these protocols have a
fixed interest rate model and cannot dynamically adjust to changing market
conditions or different borrower profiles. As a result, these protocols must be
conservative in choosing loan parameters to mitigate default risk.

\paragraph{Curatorial lending protocols.} 
Newer lending protocols, such as Morpho~\cite{metamorpho} and
Euler~\cite{euler}, introduce a third-party known as a \emph{curator}. In these
markets, curators accumulate assets from suppliers into `vaults' which each
target a particular borrower profile and, as a result, don't need such
conservative risk parameters. Competition amongst curators results in dynamic
price discrimination on the rates charged to the different
markets~\cite{bertucci2024agents}. These curatorial markets have grown to
over~\$6 billion in assets in 2025~\cite{defillama-morpho}.

\paragraph{Market clearing model.}
Consider a curatorial market with $V$ vaults (items) and $B$ borrowers (buyers).
Loans from each vault $v = 1, \dots, V$ can only be used for particular,
specified actions. Each borrower $b = 1, \dots, B$, set $(c_b)_v = 1$ if they
can use vault $v$ for their desired actions and $(c_b)_v = 0$ otherwise. If a
vault $v$ has supply $s_v > 0$ current demand $d_v \ge 0$, then for a loan of
size $x$ the vault charges an interest rate
\[
    \kappa_v \cdot \frac{d_v + x}{s_v},
\]
where $\kappa_v > 0$ is a constant. Borrower $b$ earns revenue $r_b(x_b)$, which
is a concave, nondecreasing function of the loan size $x_b$. Thus, after loans
are allocated, this buyer has net utility
\[
    u_b(x_b) = r_b(x_b) - \kappa_v \cdot \frac{d_v + x_b}{s_v}\cdot x_b.
\]
The market clearing problem aims to allocate the supply to borrowers that
minimizes their borrowing cost while honoring vault constraints.

\paragraph{Discussion.}
We could clear the other side of this market---allocation from asset suppliers
to curated vaults---with our market clearing mechanism as well. Each asset
supplier attempts to maximize their revenue while minimizing risk from bad
loans, expressed via preferences over the vaults to which they would supply
assets. Since the mechanism only requires specification of utility functions and
the preference vector, a protocol could also dynamically rebalance vaults on
behalf of the suppliers. Also note that, in practice, the interest rate
typically depends on not only an individual borrower's loan but also the loans
allocated to all other borrowers at the same time. Modeling this results in a
non-separable utility function, similar to that of our next example.

\subsection{Restaking networks}
Restaking markets in proof-of-stake cryptocurrency networks exhibit similar
behavior. These markets allows asset holders to lend out their assets for 
additional yield.

\paragraph{Staking.}
Proof-of-stake cryptocurrency networks provide economic incentives for users,
called \emph{stakers}, to validate the state of the network. Specifically, these
users earn rewards for checking the validity of blocks submitted to the network.
To earn these rewards, stakers must lock up assets as collateral. If a staker
incorrectly performs their validation duties, they are financially penalized
(`slashed', in industry parlance). See~\cite{saleh2021blockchain} for additional
details.

\paragraph{Restaking.}
Restaking allows for a staker to earn rewards for validating multiple networks
(`services', in industry parlance)~\cite{team2024eigenlayer} using the same
assets. In other words, a single staker pledges their assets as collateral to
multiple services. These stakers now have many different validation
duties---some from each service they validate---and, as a result, may be
penalized by any of these services, raising the risk of default and of
liquidation cascades.  Durvasula and Roughgarden~\cite{durvasula2024robust}
first formalized the excess default risk, and follow-up work by Chitra and
Pai~\cite{chitra2024much} analyzed the impact of how incentives paid affected
this default risk. The restaking service Eigenlayer on the Ethereum blockchain
has over \$15 billion of assets~\cite{defillama-eigenlayer}.

\paragraph{Market clearing model.}
We view each service's reward (yield) as an item and each staker as a buyer.
Each buyer aims to maximize expected revenue within their risk parameters. They
will evaluate services based on a number of properties such as historical
economic performance, penalty (slashing) rates, service uptime, current yield,
and service credit risk. In practice, these properties may be collapsed into
only rating and yield, as in~\S\ref{sec:simple-ex}. We note that restaking
allows for a single staker to delegate their stake to multiple services. We
adapt the incentivized restaking model of~\cite{chitra2024much}. Each service $s
= 1, \dots, S$ costs $c_s$ and pays rewards $r_s$, split pro-rata amongst all
its validators. We consider $SB$ items, one for each service-validator
combination. The preferences can be encoded as
\[
    (c_b)_s = \begin{cases}
        1     & \text{staker $b$ accepts service $s$} \\
        0       & \text{otherwise,}
    \end{cases}
\]
where $c_b$ only `hits' the relevant items for validator $b$. The (nonseparable,
concave) utility is then
\[
    U(x) = \sum_{s = 1}^S \left(r_s \cdot \frac{x_{b,s}}{\sum_{b=1}^B x_{b,s}}
    - c_s \cdot {\textstyle \sum_{b=1}^B x_{b,s}}\right).
\]
We can similarly clear this market by solving~\eqref{eq:market-clearing}. Note
that, while the objective is not separable, we conjecture that can use the same
ideas outlined in~\S\ref{sec:dsic}: charging buyer $b$ her externality results
in a DSIC payment rule. We leave full characterization to future work.

\section{Conclusion}
In this paper, we constructed markets over semi-fungible (or
partially-ordered) assets. Such markets take into account this partial ordering
in order to provide additional liquidity for buyers and sellers:
any buyer who wishes to buy an item is, if they receive an item, guaranteed an
item at least as good as the requested one. Thus, any
seller has `more chances' to sell their items. We then showed that clearing
these markets is a convex optimization problem that can be efficiently solved
on modern hardware. The dual problem admits a nice interpretation in
terms of the market clearing prices, and we showed that these prices respect the
partial order; \ie, items that are `better' with respect to this order have a
price no lower than those which are `worse'. Finally, we constructed a dominant
strategy incentive compatible mechanism such that buyers are incentivized to
bid their true valuations, which is efficient to implement in practice. We
believe that our mechanism opens many avenues for interesting future work.


\section*{Acknowledgments}
The authors thank Ryan Cohen for initially inspiring this line of research and
for many helpful discussions.

\ifsubmit
    \bibliographystyle{ACM-Reference-Format}
    \bibliography{refs}
\else
    \printbibliography
\fi

\end{document}